\documentclass{llncs}
\usepackage{amsmath}
\usepackage{amssymb}
\usepackage{z-eves}
\usepackage{xspace}
\usepackage{soul}
\usepackage[dvipsnames*,svgnames]{xcolor}

\usepackage{tikz}
\usetikzlibrary{decorations.text, positioning}
\usetikzlibrary{arrows.meta}

\newcommand{\step}[1]{\mathbin{\lower0.55ex\hbox{$\lhook\joinrel\xrightarrow{#1}$}}}

\newlength{\bcextramargin}
\newenvironment{changemargin}[2]{\begin{list}{}{%
\setlength{\topsep}{0pt}%
\setlength{\leftmargin}{0pt}%
\setlength{\rightmargin}{0pt}%
\setlength{\listparindent}{\parindent}%
\setlength{\itemindent}{\parindent}%
\setlength{\parsep}{0pt plus 1pt}%
\addtolength{\leftmargin}{#1}%
\addtolength{\rightmargin}{#2}%
}\item }{\end{list}} 

\newcommand{\actdefsection}[1]{
\begin{changemargin}{-\bcextramargin}{0pt}
\vspace{1ex}
\noindent
\textbf{{#1}}
\end{changemargin}
}

\newenvironment{absolutelynopagebreak}
  {\par\nobreak\vfil\penalty0\vfilneg
   \vtop\bgroup}
  {\par\xdef\tpd{\the\prevdepth}\egroup
   \prevdepth=\tpd}

\newcommand{\setlog}{$\{log\}$\xspace}

\newcommand{\bs}{\boldsymbol{\sigma}}

\begin{document}
\bibliographystyle{plain}
\title{Set-Based Models for Cryptocurrency Software}

 \author{Gustavo Betarte\inst{1} \and Maximiliano Cristi\'a\inst{2} \and Carlos Luna\inst{1} \and Adri\'an Silveira\inst{1} \and Dante Zanarini\inst{2}} 
 \institute{InCo, Facultad de Ingenier\'ia, Universidad de la Rep\'ublica, Uruguay. 
    \\ \email{\{gustun,cluna,adrians\}@fing.edu.uy} 
 \and CIFASIS, Universidad Nacional de Rosario, Argentina.
  \\ \email{\{cristia,zanarini\}@cifasis-conicet.gov.ar}
 }

\maketitle

\begin{abstract}                                          
Emin G\"un Sirer once said\footnote{\url{hackingdistributed.com/2016/06/17/thoughts-on-the-dao-hack}.}: \emph{It's clear that writing a robust, secure smart contract requires extreme amounts
of diligence. It's more similar to writing code for a nuclear power reactor, than to
writing loose web code [\dots] Yet the current Solidity language and underlying EVM seems designed more
for the latter.}

Formal methods (FM) are mathematics-based software development methods aimed at producing ``code for a nuclear power reactor''. That is, due application of FM can produce bug-free, zero-defect, correct-by-construction, guaranteed, certified software. However, the software industry seldom use FM. One of the main reasons for such a situation is that there exists the perception (which might well be a fact) that FM increase software costs. On the other hand, FM can be partially applied thus producing high-quality software, although not necessarily bug-free.

In this paper we outline some FM related techniques whose application the cryptocurrency community should take into consideration because they could bridge the gap between ``loose web code'' and ``code for a nuclear power reactor''. 

\end{abstract}
\section{Introduction}
\label{sec:intro}

Given that cryptocurrency software deals with virtual money, software errors can produce irreparable loses. Furthermore, they are a valuable target of highly skilled attackers. Therefore, if a cryptocurrency software has an exploitable vulnerability chances are that attackers will eventually use it to steal money. In fact, some attacks have already been mounted against cryptocurrency software causing irreparable loses of money and credibility (e.g. \cite{dao}). 

Note that cryptocurrency software face a more complex problem than traditional banking systems. If your homebanking account is hacked your money can be (potentially) wired transferred to one or more \emph{bank accounts} anywhere on Earth. But these accounts have registered owners who can be duly prosecuted, although some times you will not get your money back. If the same happens in the cryptocurrency world you will not have a registered owner and so prosecution will be harder, if possible.

Hence, software errors in the cryptocurrency world are potentially more costly than the same class of errors in traditional banking systems. Therefore, the quality of cryptocurrency software should be, at least, one ``level'' higher than that of banking software. Although banking software is not (always) ``loose web code'', it certainly is not ``code for a nuclear power reactor''.

For these reasons the cryptocurrency community is seeking for approaches, methods, techniques and development practices that can reduce the chances of the presence of either errors or vulnerabilities\footnote{Actually, the presence of certain errors lead to the existence of vulnerabilities.}. The traditional banking system has less incentives to pursue high levels of software quality and thus to seek better development practices.

One such approach is the application of Formal Methods (FM) to software construction. FM are development methods based on mathematics and logic. They have been around in academia for at least 50 years. There have been undergraduate courses around the globe teaching FM for decades. A number of scientific journals and international conferences are devoted to the study, progess and application of FM. FM is the working field of hundreds of researchers across the planet. The FM community has produced breakthrough results on fundamental aspects of Computer Science and some of its most prominent actors have earned international prizes such as the Turing Award. Leading companies such as Microsoft and Amazon hire  software engineers or researchers on FM to produce or apply FM.

When FM are fully applied they produce ``code for a nuclear power reactor''. That is, due application of FM can produce bug-free, zero-defect, correct-by-construction, guaranteed, certified software. There are a number of critical or mission critical software systems that were developed with FM. 

However, the software industry seldom use FM. There are several reasons for which industry is reluctant to use FM among which we can cite: a pervasive underrating of software quality (usually in favor of innovation or as the consequence of free distribution\footnote{That is: Why should I produce bug-free software if I am not sure anyone will use it? Why are you asking me for a warranty if I am giving you this program for free?}); either the perception or the fact that FM severely increase costs and extend schedules; at times the sheer fact that managers do not have a clue about the existence of FM; and the difficulties of finding the right technical staff. 

In this paper we present the case for the application of FM to cryptocurrency software. In Sect. \ref{case} we present some guidelines for the adoption of FM in cryptocurrency software projects. We argue that set-based formal modeling (or specification), simulation, prototyping and automated proof can be applied before considering more powerful approaches such as code formal verification. Hence, in Sect. \ref{protocol} we show excerpts of a set-based formal specification of a consensus protocol and in Sect. \ref{evm} of the Ethereum Virtual Machine (EVM). In Sect. \ref{verification} we show that prototypes can be generated from these formal models and simulations can be run on them. Then, we show that test cases can be generated from the same models and how automated proof can be used to evaluate the correctness of these models.

\section{\label{case}Guidelines for Formal Methods Adoption}
This section presents guidelines that the cryptocurrency community can consider for the adoption of FM. It starts with a brief and broad presentation of FM. 

\subsection{Formal methods in a nutshell}
As we have said, FM is a class of software development techniques based on mathematics and formal logic that can be applied to different  development activities and phases\footnote{FM can be applied to hardware systems, too.}. The main goal of FM is to improve the confidence on the correctness of a program. Program correctness is usually defined as follows: \emph{a program is correct if it verifies its specification}. The word `verifies' can be replaced by `satisfies', `refines', `respects', etc. In any case, the fundamental idea is that you have a program, $P$, and its specification, $S$, and a FM provides you a way to ensure, with different levels of confidence, that $P$ behaves as $S$ states. 

In FM the specification or model, $S$ plays a central role:
\begin{itemize}
\item $S$ should be written from the requirements, not from the program; ideally $S$ should be written before $P$ is developed.
\item $S$ is a formal description; that is, in a sense, $S$ is a mathematical model or formula.
\item $S$ represents a family of programs: from $S$ you can get a number of different programs; $S$ does not fix a particular implementation.
\item $S$ should be simpler and shorter than $P$ but, fundamentally, it should be more evidently correct than $P$. That is, when you read $S$ you should feel you are reading something that obviously describes what you want.
\item $S$ should abstract away as many implementation details as possible. Implementing a program entails giving enough details as to a  computer can execute it. A huge number of these details are nonexistent in the requirements. For example, at the requirements level one simply says ``if the customer is a new one, then add it to the data base'', while a C implementation might use a singly-linked list to store customers and so checking if the customer exists and adding it requires pointer arithmetic. In this case, pointer arithmetic is an implementation detail that should not be visible in $S$.
\item $S$ is usually a functional specification but it can also be a security specification (or whatever other non-functional requirement). Then, FM can be used to verify that $P$ correctly does what it is supposed to do and that it does that securely. 
\end{itemize}

Then, the application of FM starts by writing a formal specification of your software. It does not mean you must write a complete specification; it can be the specification of a portion of your program (usually the most complex or the most critical). It has been proved countless times that the mere fact of writing $S$ clarifies ideas, pops up unseen problems, etc. Is this a waste of time and resources, or an added cost? No, it is not. These unseen problems will flow into code and sooner or later they will produce unwanted behaviors. Actually, writing $S$ can save you time and money.

Depending on the particular FM you are using, once you have $S$ you can keep applying formal techniques such as:
\begin{enumerate}
\item $S$ can be checked for different levels of soundness.
\item $S$ can be used as a prototype.
\item $S$ can be used to generate test cases.
\item $S$ can be handled to the development team (i.e $P$ is implemented from $S$).
\item $S$ can be used to formally generate a correct-by-construction $P$.
\end{enumerate}

All these activities improve the confidence on the correctness (either functional or security) of your software; only the last one can guarantee correctness but it is the most demanding in terms of time, effort and technical proficiency. Depending on the context, many of these techniques can be automated to some extent, and some times they can be fully automated. The first activity can uncover errors in $S$ thus removing them before they make into $P$. The difference between 4 and 5 is that in 4 developers just read $S$ and write $P$ while in 5 developers \emph{formally prove} that $P$ implements $S$.

There are a number of FM and formal techniques. There are old, enduring, established, assessed, supported FM and there are new, experimental ones. None of them is a silver bullet: if it is very expressive, then it will  be harder to automate, and vice versa; if its models are abstract, then program correctness will be harder to be formally proved; if its models are less abstract, then their complexity can be similar to that of the implementation. In general, each FM was thought and designed to be applied to a class of systems or for a particular verification activity. For example, there are FM to specify and analyze information systems; others more suitable for concurrency; there are very expressive FM aimed at full formal verification; and others aimed at automated verification of limited classes of programs.

In any case, the determining factor of a FM is its specification language. The specification language somewhat fixes what can be expressed, what formal techniques are possible and which can be automated to some valuable extent. The specification language determines the form of your models or specifications. Each specification language is based on some well-known mathematical or logic theory; they always have a formal semantics. There are specification languages based on set theory, type theory, relation algebra, process algebra, transition systems, first-order logic, category theory, higher-order logic, temporal logic, etc. Some specification languages resemble high-level programming languages, most look like math+logic. Nevertheless, specification languages are to high-level programming languages what the latter are to machine code. That is, you can implement a formal specification in any given programming language.

\subsection{Components of cryptocurrency software}
A cryptocurrency software may encompass the following: cryptographic primitives, cryptographic protocols, consensus protocols, a blockchain, a virtual machine, a scripting language, a contract-oriented programming language, a wallet a number of smart contracts. Orthogonal to those the system runs on top of one or more operating systems and the TCP/IP stack, and it has been programmed in one or more programming languages with their corresponding compilers or interpreters. Yet orthogonal to all that you have complex functional and security requirements. And finally, a cryptocurrency software is usually a distributed system. 

Any error on any of those components can potentially produce a failure or a vulnerability. That is, an error on any of those components can potentially affect the functional correctness of the system or its security.

Guaranteeing that a cryptocurrency software will never fail and will never be hacked (at least by using any known attack technique) entails proving that all those components are functional and security correct, which in turn requires a formal specification of each component and the interactions between them---which, in particular, includes giving a formal semantics to all the programming languages involved. If you program your blockchain on Rust and the Rust compiler has some error, then your system is potentially vulnerable; if you are using a proven cryptographic primitive whose implementation happens to hide an obscure error, then your system is potentially vulnerable.

Certifying the functional and security correctness of such a system is, currently, a challenging task even for seasoned FM experts. Fortunately, in many cases each component can be independently certified and, furthermore, many verification tasks (cf. items 1-4 above) can be carried on before attempting a formal correctness proof. Indeed, so far, the FM community has formally verified (some times parts of) some of the components of a cryptocurrency software such as, cryptographic primitives \cite{DBLP:journals/iacr/ProtzenkoPFHPBB19}, compilers \cite{DBLP:journals/cacm/Leroy09}, communication protocols \cite{DBLP:conf/nsdi/MusuvathiE04}, cryptographic protocols \cite{DBLP:conf/csfw/BartheHBGH10}, secure operating systems \cite{DBLP:journals/tocs/KleinAEMSKH14}. However, as far as we know, a formal proof of the scale and scope necessary for a full-fledged cryptocurrency software has never been attempted.

\subsection{Cryptocurrency software as critical systems}
If society will eventually depend on a cryptocurrency software its correctness proof should be attempted. Indeed, if a cryptocurrency software ever plays an important role in the society it will be that of a critical system. That is, of a system whose failure would cause irreparable loses---as a failure in the software controlling a nuclear power reactor. If the error or the attack causes the unwanted transfer of cryptomoney, the victims will not be able to recover it due to the virtual nature of cryptocurrencies and because many of these systems enforce strong privacy and anonymity protections. 

On the other hand, critical software has been the main target of FM since their inception. For example, the railway industry has been applying FM for its critical systems for quite a long time \cite{DBLP:conf/sbmf/LecomteDPM17}; computer security has been one of the traditional application domains of FM for almost 50 years \cite{Bell:LaPadula,GoguenM82,DBLP:journals/tocs/KleinAEMSKH14,Barthe2017,DBLP:journals/cuza/BetarteCLR16,DBLP:conf/ccs/BartheBCLP14}; several other industrial sectors such as nuclear, defense, health care, etc. apply FM to some of their critical software projects where, in average, 80\% report no increase in development time, +90\% report no added costs, and 88\% report on improved quality \cite{DBLP:books/daglib/p/FitzgeraldBLW13}. 

Developers of cryptocurrency software should not be scared about using mathematics as a tool to describe software. In fact, Nakamoto uses math in his seminal paper on Bitcoin \cite{bitcoin} and Wood uses it to describe the EVM \cite{wood2014ethereum}. However, these descriptions would not be understood as FM because they are not based on standardized notations nor on clear mathematical theories. On the other hand, recently, the FM community has started to pay attention to cryptocurrency software. Idelberger et al. \cite{DBLP:conf/ruleml/IdelbergerGRS16} propose to use defeasible logic frameworks such as Formal Contract Logic for the description of smart contracts. Bhargavan et al. \cite{DBLP:conf/ccs/BhargavanDFGGKK16} compile \textsc{Solidity} programs into a verification-oriented functional language where they can verify source code. Luu et al. \cite{DBLP:conf/ccs/LuuCOSH16} use the \textsc{Oyente} tool to find and detected vulnerabilities in smart contracts. Hirai \cite{DBLP:conf/fc/Hirai17} use \textsc{Lem} to formally specify the EVM; Grishchenko, Maffei and Schneidewind \cite{DBLP:conf/post/GrishchenkoMS18} also formalize the EVM but in \textsf{F*}; and Hildenbrandt et al. do the same but with the reachability logic system known as $\mathbb{K}$. P\^irlea and Sergey \cite{Pirlea:2018:MBC:3176245.3167086} present a Coq \cite{coq-manual,coqart} formalization of a blockchain consensus protocol where some properties are formally verified. 
Finally, in \cite{DBLP:journals/corr/abs-1907-01688} Betarte et al. present and briefly discuss (formally) properties and outline the basis of a model-driven verification approach to address the certification of the correctness of a particular implementation of the MimbleWimble cryptocurrency protocol.

\subsection{A gradual adoption process}
Hence, the cryptocurrency community should consider the adoption of FM for the developtment of its critical software components. However, the adoption of FM is not only a technical challenge but, more importantly, it requires a sort of cultural shift. In effect, the introduction of FM into the development process requires new personnel or a thorough training of existing staff; the introduction of new tools, techniques and processes; and, in some cases, a redesign of the development cycle---e.g., you cannot change source code without first (re)proving a theorem. Note that the works cited above are carried out by experts on FM. If the cryptocurrency community wants to use FM then it either has to establish strong, mid-term collaboration projects with academic groups or start an adoption program.

Due to the nature and dynamics of the cryptocurency ecosystem we think that developers should slowly introduce FM into the development of cryptocurrency software. FM should become a new asset of the community rather than an outsourced service. The following are some guidelines that can be considered:
\begin{enumerate}
\item Do not attempt a full formal correctness proof from the beginning.
\item Start with abstract specification languages (e.g. set theory) based on first-order logic.
\item Learn to formally specify.
\item Put the verification of code aside for later, focus on writing good models.
\item Abstract away cryptography, assume it is correct, focus on functional specifications.
\item Add \emph{lightweight} or \emph{automated} verification techniques such as simulation \cite{DBLP:journals/scp/HallerstedeLP13}, prototyping \cite{CristiaRossiSETS14}, model-based testing \cite{Utting00}, model-checking \cite{DBLP:books/daglib/0007403} and automated proofs \cite{DBLP:books/daglib/0007471,DBLP:series/faia/2009-185}.
\item Assess what you have learned, correct the course and move on. 
\end{enumerate}
 
In point 5 we suggest that cryptography should be abstracted away in this first adoption stage. This might be generalized to security properties. 
The formal verification of security properties is still a challenging issue (see Appendix \ref{security}). Formal verification of some cryptographic primitives has only been achieved very recently \cite{DBLP:journals/iacr/ProtzenkoPFHPBB19}. Furthermore, formally proving non-trivial security properties of code might be an overwhelming task in terms of the effort required, especially compared with proving functional correctness. In addition, many implementation details are orthogonal to the security properties to be established. This implies that slight changes in the implementation technology might have devastating consequences as concerns the security of the implementation.

Therefore, in this paper we show excerpts of a set-based formal model of two components of a cryptocurrency software: a consensus protocol (Sect. \ref{protocol}) and the Ethereum Virtual Machine (EVM, Sec. \ref{evm}), where cryptography and security are abstracted away. Besides, we show how these models can be easily, although partially, analyzed by either lightweight or automated verification techniques (Sect. \ref{verification}).

\section{\label{protocol}Formal Specification of a Consensus Protocol}
There are several FM based on set theory and first-order logic (e.g. Z \cite{Spivey00}, VDM \cite{Jones:1990:SSD:94062}, B \cite{Abrial00}). These notations are designed to write large, complex specifications; and there is a number of tools to work with them such as editors, typecheckers, theorem provers, animators, etc. However, in this paper we will use the plain and simple language of mathematics extended with a couple of conventions, to avoid explaining the peculiarities of a particular notation. This will be enough as to show the `look-and-feel' of such specifications.

The following are some snippets of a model of a consensus protocol based on the work by P\^irlea and Sergey \cite{Pirlea:2018:MBC:3176245.3167086}. 

A consensus protocol deals with addresses, hashes, proof objects, etc. In an abstract model of the protocol the internal structure of these entities is irrelevant. In particular, hashes and proof objects are strongly related to cryptography which is a feature we think should be abstracted away at this point. Then, we have the set of all possible addresses ($Addr$) that can be used. If $a \in Addr$ then $a$ is an address of the protocol. We do not know what $a$'s structure is, how it was generated, whether it is 128 bits or 256 bits long, etc. The idea is that we do not need to know those things now, because they are implementation details which will not alter the fundamental behavior of the protocol. In this sense we also have the set of hashes ($Hash$), the set of proofs objects ($Proof$) and the set of transactions ($Tx$). The only condition required for these sets is that they come equipped with equality and be pairwise disjoint. 
\[
[[Addr,Hash,Proof,Tx]]
\]
The block data structure (cf. blockchain) is a record with three fields: $prev$, (usually) points to the parent block; $txs$, stores the sequence of transactions stored in the block; and $pf$ is a proof object required to validate the block. Then we define $Block$ as the set of all such records:
\[
Block \defs [prev:Hash; txs: \seq Tx; pf:Proof]
\]
The local state space of a participating network node is given by three state variables: $as$, are the addresses of the peers this node is aware of; $bf$, is a block forest (not shown) which records the minted and received blocks; and $tp$, is a set of received transactions which eventually will be included in minted blocks.
\[
LocState \defs [as:\power Addr; bf: Hash \pfun Block; tp: \power Tx]
\]
The protocol configuration is represented by two state variables: $Delta$, which establishes a mapping between network addresses and the corresponding node (local) states (in \cite{Pirlea:2018:MBC:3176245.3167086} this variable is referred to as the \emph{global state}); and $P$, a set of packets (which represent the messages exchanged by nodes). 
\[
Conf \defs [Delta:Addr \pfun LocState; P:\power Packet]
\]
Packets are just tuples of two addresses (origin and destination) and a message.
\begin{zed}
Packet == Addr \cross Addr \cross Msg
\end{zed}
The model has twelve state transitions divided into two groups: \emph{local} and \emph{global}. Local transitions are those executed by network nodes, while global transitions promote local transitions to the network level. In turn, the local transitions are grouped into \emph{receiving} and \emph{internal} transitions. Receiving transitions model the nodes receiving messages from other nodes and, possibly, sending out new messages; internal transitions model the execution of instructions run by each node when some local condition is met. Here, we show only the local, receiving transition named $RcvAddr$.
\[
RcvAddr(s:LocState; p?:Packet; ps!:\power Packet; s':LocState) \defs \\
\t1 p?.2 = this \\
\t1 \land \exists asm:\power Addr @ \\
   \t2 p?.3 = AddrMsg~asm \\
   \t2 \land s'.as = s.as \cup asm \\
   \t2 \land s'.bf = s.bf \land s'.tp = s.tp \\
   \t2 \land ps! = \{a:asm \setminus as @ (p?.2,a,ConnectMsg)\} \\
   \t4{}\cup \{a:as @ (p?.2,a,AddrMsg~p'.as)\}
\]
As can be seen, $RcvAddr$ transitions from the local state $s$ into a new local state $s'$, receives a packet ($p?$), and sends out a set of packets ($ps!$). The node checks whether or not the packet's destination address coincides with its own address. In that case, the node adds the received addresses to its local state and sends out a set of packets that are either of the form $(p?.2,a,ConnectMsg)$ or $(p?.2,a,AddrMsg~p'.as)$. The former are packets generated from the received addresses and sent to the new peers the node now knows, while the latter are messages telling its already known peers that it has learned of new peers.

Observe that the expression after the $\defs$ symbol in $RcvAddr$ is a predicate. That is, for example, the equal symbol in $s'.as = s.as \cup asm$ represents logical equality and not (imperative) assignment; $\land$ means logical conjunction and thus $p \land q$ is equal to $q \land p$; $\cup$ is set union; and so forth. 

Hopefully, this simple example lets the reader see that abstract specifications tend to be much simpler, concise and evidently correct than code. Furthermore, much of what is said in an abstract specification is what seasoned programmers think while designing good code. Then, formal specification languages give programmers an efficient technique by means of which they can write down their best ideas and be able to communicate them easily and unambiguously. In most cases using formal specifications in this way provides costs savings rather than the opposite.

\section{\label{evm}Formal Specification of the EVM}
The same notation and methodology can be used to formally specify the EVM. In this case we depart from the Yellow Paper \cite{wood2014ethereum}. The state of the EVM is given by three records:
\[
EVMState \defs [World; Machine; CallStack]
\]
where, for example $World$ and $Machine$ are defined as follows:
\[
World \defs [acc,accCC:Addr \pfun Acc; newaddr:Addr; step:STEP] \\
Machine \defs [
g:ETH;
pc: \nat;
m: A \pfun B;
i:\nat;
s: \seq W;
out: P]
\]
where in turn, for instance, $acc$ represents the Ethereum accounts as a partial function from the set of addresses onto the set of records $Acc \defs [nonce:\nat; bal:ETH; code: PROG]$, which stores the main information of the accounts; $step$ records the current transaction execution step; $m$ represents the state of the memory and $g$ the available gas---where $A$, $B$, $W$, etc. are given sets like $Addr$. 

We have two types of transactions: 
\[
TT \defs \{contractCreation, messageCall\}
\]
A transaction can be modeled as a record with several fields (some are omited):
\[
Transaction \defs
[Tn,Tg:\nat;
Tp,Tv: ETH;
Ti: P;
Td: \seq B;
snd:Addr]
\]
where, for instance, $Tn$ is the nonce, $Tg$ is the gas limit, $Tp$ is the gas price and $Ti$ is the initialization program for the account.

Now we can model how transactions are executed. According to the YP, a transaction is executed in four steps:
\emph{(a)} the checkpoint state ($\bs_0$, in YP's notation); \emph{(b)} the post-execution provisional state ($\bs_P$); \emph{(c)} the pre-final state ($\bs^*$); and \emph{(d)} the final state ($\bs'$) is reached after deleting all accounts
that either appear in the self-destruct list or are touched
and empty. The specification of the processing step called checkpoint state ($\bs_0$) is independent of the transaction type and is defined as follows:
\[
CheckpointState(s:World; t?:Transaction; s':World) \defs \\
\t1 TransactionValidity(s,t?) \\
\t1 \land UpdateSender(s.acc~t?.sender, (acc~t?.sender).bal, t?.Tp, t?.Tg,a')\\
\t1 \land s.step = initial \\
\t1 \land s'.acc = s.acc \oplus \{(t?.sender,a')\} \\
\t1 \land s'.step = ccbegins
\]
where $TransactionValidity$ (not shown) is a complex predicate stating the conditions for a transaction to be valid and $UpdateSender$ is defined as follows:
\[
UpdateSender(s:Acc;b,p:ETH;g:\nat;s':Acc) \defs \\
\t1 s'.bal = b - g*p
\land s'.nonce = s.nonce + 1
\land s'.code = s.code
\]

Hence, $CheckpointState$ checks whether the requested transaction is valid and in that case it (only) updates the sender account by debiting the result of multiplying the gas price and the gas limit as given in the transaction and by adding one to the number of transactions ($s.nonce$).

The transition from $\bs_0$ to $\bs_P$ requires the execution of the program associated to the account (or the initialization program if it is a contract creation). Then, at this point we should specify how the EVM executes (bytecode) programs. 
To this end, we first need to formalize each and every EVM bytecode instruction. Most instructions can be modeled as a transition between two $MachineState$. We will do that only for a simplified version of the instruction called \verb+create+. According to the YP \verb+create+ ``Creates a new account with associated code''. We have identified two cases so we have:
\begin{zed}
Create \defs Create1 \lor Create2
\end{zed}
where $Create1$ is not shown due to space restrictions. $Create2$ formalizes a situation where the account is actually not created due to some boundary condition:
\[
Create2(q:Machine; w:World; a?:Addr; n?:\nat; q':Machine) \defs \\
\t1 q.s~1 > (w.acc~a?).bal \lor n? \geq 1024 \\
\t1 \land q'.s = \langle 0 \rangle \cat tail(tail(tail~q.s)) \\
\t1 \land q'.i = M(q.i,q.s~2,q.s~3) \\
\t1 \land q'.m = q.m \land q'.g = q.g 
\]
where $M$ updates the amount of used memory ($q.i$) by using the second and third positions of the machine stack ($q.s$). Note that $Create2$ modifies the machine state but it needs to access the world state. It also receives the address of the account which owns the code
that is executing ($a?$) and the number of $Creates$ being executed at present ($n?$). The account is not created because either the balance of the caller is too low to fulfill the value transfer or there are too many active $Create$ calls ($n? \geq 1024$). In this case the first three positions in the stack are removed and a 0 is stacked on top of that. 

\paragraph{Lesson learned.} The YP describes the semantics of the EVM mixing informal text with  formulas written in some ad-hoc mathematical notation. In the process of writing this specification we found many obscure (probably inconsistent) issues in the math used in the YP. Established formal specification languages have gone a long process of standardization and analysis; their fundamentals have been studied for decades (maybe centuries) by some of the founding fathers of modern mathematics and Computer Science. Even the math used in the YP cannot be compared to well established FM. Besides, most FM are supported by tools implementing a variety of verification techniques. Using math in the YP is a good step forward but the invested effort would be at least partially wasted.

\section{\label{verification}Some Verification Techniques for Set-Based Models}
In this section we show the application of three verification techniques that can be used when set-based formal specifications are available. In the first one we show how a set-based specification can be turned into a set-based prototype, that is an inefficient program that nonetheless is correct-by-construction and so it can be used to analyze complex situations (Sect. \ref{prototypes}). Next, we show how the same set-based model can be used to generate test cases that can be used to test the implementation (Sect. \ref{mbt}). Finally, we automatically prove that the same models enjoy some properties (Sect. \ref{proofs}).

\subsection{\label{prototypes}Set-based prototypes and simulations}

Set-based specifications such as those presented in sections \ref{protocol} and \ref{evm}, can be easily turned into set-based prototype programs. In effect, a set-based model can be encoded in the programming language provided by the \setlog (`setlog') tool \cite{Cristia2019,setlog}. \setlog is a programming language, a satisfiability solver and an automated theorem prover where sets are first-class entities. \setlog provides the usual Boolean connectives and most of the set and relational operators available in set theory including binary relations. Hence, it is quite natural to encode a set-based specification as a \setlog program. Given that \setlog is based on Prolog its programs resemble Prolog programs. 

The \setlog encoding of $RcvAddr$ presented in Sect. \ref{protocol} is the following:
\begin{verbatim}
rcvAddr(S,P,Ps,S_) :-
  S = {[as,As] / Rest} &
  P = [_,this, addrMsg(Asm)] &
  un(As,Asm,As_) &
  diff(Asm,As,D) &
  PsD = ris(A in D,[],true,[this,A,connectMsg]) &
  PsAs = ris(A in As,[],true,[this,A,addrMsg(As_)]) &
  un(PsD,PsAs,Ps) &
  S_ = {[as,As_] / Rest}.
\end{verbatim}

\verb+rcvAddr+ is a \setlog clause respecting the interface of $RcvAddr$. In this case, however, instead of using set membership to somewhat ``type'' the parameters we rest on unification. As in Prolog, \setlog programs are based on unification with the addition of set unification. Variables must start with an uppercase letter. We take advantage of set theory to encode other data structures such as records. A record is a set of ordered pairs where the first component names the field and the second is the variable holding the values. Hence, a statement such as \verb+S = {[as,As] / Rest}+ (set) unifies the first parameter with a set term singling out the record field needed in this case (\verb+As+) and the rest of the record (\verb+Rest+). The same is done with packet \verb+P+ where \verb+_+ means any value as first component and \verb+addrMsg(Asm)+ gets the set of addresses received in the packet without introducing an existential quantifier. The set comprehensions used in the specification are implemented with \setlog's so-called Restricted Intentional Sets (RIS) \cite{DBLP:conf/cade/CristiaR17}. A RIS is interpreted as a set comprehension where the control variable ranges over a finite set (\verb+D+ and \verb+As+). Finally, \verb+diff(A,B,C)+ is interpreted as $C = A \setminus B$ and \verb+un(A,B,C)+ as $C = A \cup B$.

A clause such as \verb+rcvAddr+ can be seen both as a \setlog formula and as a \setlog program. It is a formula in the sense that \verb+&+ is logical conjunction and so the order of statements in \verb+rcvAddr+ is irrelevant due to commutativity. It is a program simply because we can use the \setlog interpreter to compute outputs from inputs. However, given that \setlog programs are meant to be prototypes we talk of \emph{simulations} or \emph{animations} rather than executions. 

Then, given that \verb+S+ and \verb+P+ are meant to be inputs while \verb+Ps+ and \verb+S_+ are outputs, we can run a simulation from an initial $LocState$ and input packet $p?$ such as:
\begin{verbatim}
S = {[as,{}] / _}
P = [_,this,addrMsg({a1,a2})]
\end{verbatim}
Now, for instance, we can call \verb+rcvAddr+ twice chaining before and after states as follows:
\begin{verbatim}
S = {[as,{}] / _} &
P = [_,this,addrMsg({a1,a2})] &
rcvAddr(S,P,Ps1,S1) & 
rcvAddr(S1,[_,this,addrMsg({a1,a3})],Ps2,S2).
\end{verbatim}
in which case \setlog returns:
\begin{verbatim}
Ps1 = ris(A in {a1,a2/_N2},[],true,[this,A,connectMsg],true),  
S1  = {[as,{a1,a2}]/R},  
Ps2 = {[this,a3,connectMsg],[this,a1,addrMsg({a2,a1,a3})],
      [this,a2,addrMsg({a2,a1,a3})] /
      ris(A in _N1,[],true,[this,A,connectMsg],true)},  
S2  = {[as,{a2,a1,a3}]/R}
Constraint: subset(_N2,{a1,a2}), subset(_N1,{a1,a3}), 
            a1 nin _N1, a2 nin _N1
\end{verbatim}
That is, \setlog binds values for all the free variables in a way that the formula is satisfied (if it is satisfiable at all). In this way we can trace the execution of the protocol w.r.t. states and outputs by starting from a given state (e.g. \verb+S+) and input values (e.g. \verb+[_,this,addrMsg({a1,a2})]+), and chaining states throughout the execution of the state transitions included in the simulation (e.g. \verb+S1+ and \verb+S2+).

\subsection{\label{mbt}Model-based testing}
Model-based testing (MBT) is a testing methodology where test cases are drawn from models or program specifications \cite{Utting00}. That is, instead of letting testers to think what test cases are necessary to test a program, a MBT method gives a disciplined, systematic and quantifiable algorithm for test case generation. There are a number of MBT methods depending on the FM and the type of systems under consideration \cite{Hierons02-largo}. Most MBT methods are supported by (semi)automatic tools thus turning program testing into a more efficient process. 

In particular, the Test Template Framework (TTF) can be applied to set-based specifications \cite{Stocks2}. Then, this means that, by writing a set-based specification you can get, almost for free, a prototype and test cases to test the implementation.

We will apply a reduced version of the TTF to the $CheckpointState$ specification given in Sect. \ref{evm}---see elsewhere in the literature for more examples and applications to critical software \cite{DBLP:conf/sew/CristiaAFPM11,DBLP:journals/stvr/CristiaAFPM14}. Note that in $CheckpointState$ we have the following post-condition:
\begin{equation}\label{eq:oplus}
s'.acc = s.acc \oplus \{(t?.sender,a')\}
\end{equation}
where $\oplus$ is a relational operator stating that if $t?.sender \in \dom s.acc$ then its relational image must be updated with $a'$, otherwise the pair $(t?.sender,a')$ must be added to $s.acc$. Besides, recall that $acc$ is a partial function, i.e. a binary relation where all first components are different from each other. Such data structures and operators are seldom available in high-level programming languages. Hence, an efficient implementation of a predicate like \eqref{eq:oplus} will yield a non-trivial piece of code that deserves to be thoroughly tested, specially when it belongs to a critical system.

The TTF defines so-called \emph{standard partitions} for each set theoretic operator. The standard partition for $R \oplus G$ is the following:
\begin{center}
\begin{tabular}{ll}
$R = \emptyset, G = \emptyset$ & $R \neq \emptyset, G \neq \emptyset, \dom G \subset \dom R$ \\[1mm]
$R = \emptyset, G \neq \emptyset$ & $R \neq \emptyset, G \neq \emptyset, \dom R \cap \dom G = \emptyset$ \\[1mm]
$R \neq \emptyset, G = \emptyset$ & $R \neq \emptyset, G \neq \emptyset, \dom R \subset \dom G$ \\[1mm]
$R \neq \emptyset, G \neq \emptyset,$ & $R \neq \emptyset, G \neq \emptyset, \dom R \cap \dom G \neq \emptyset,$ \\
$\quad \dom R = \dom G \phantom{a}$ & $\quad \lnot(\dom G \subseteq \dom R), \lnot(\dom R \subseteq \dom G)$
\end{tabular}
\end{center}
This means that an expression of the form $R \oplus G$ should be tested with eight test cases as follows: $R$ and $G$ equal to the empty set; $R$ equal to the empty set and $G$ not equal to the empty set and vice versa; $R$ and $G$ different from the empty set but having the same domain; and so forth. If for a particular expression some combination of $R$ and $G$ is unfeasible, then it is simply discarded.

Thus, when the standard partition for $\oplus$ is applied to \eqref{eq:oplus} in the context of $CheckpointState$ it yields, after some simplifications, only two test conditions:
\[
\dom s.acc = \{t?.sender\} \\
\{t?.sender\} \subset \dom s.acc
\]
That is, it make sense to test the implementation of \eqref{eq:oplus} in $CheckpointState$ when: the address of the transaction sender is the only account in the system and when it is not the only account in the system.
However, there is more to be tested in $CheckpointState$. For example, $TransactionValidity$ is a complex predicate where the TTF can be applied in several ways. Furthermore, the TTF dictates how the test conditions for some lines of code must be combined with the test conditions generated for the other lines of code.

In general, all the steps of the TTF can be (semi)automated by tools such as \textsc{Fastest} \cite{DBLP:journals/stvr/CristiaAFPM14} and other MBT methods for set-based specifications provide good tool support as well \cite{DBLP:journals/sttt/LeuschelB08}.

\subsection{\label{proofs}Automated proofs}
A program is correct if it verifies its specification. But if the specification is wrong, the program will be wrong and this will be undetectable. Therefore, effort must be made to ensure the specification is correct. The confidence on the correctness of the specification can be increased by proving that it enjoys many desired properties. In a full formal context, all these proofs should be mechanized, although a good first approximation are manual proofs and a second best are automated proofs, as in general full automation is impossible. 

For set-based specifications \setlog can be used as an automated theorem prover and as a counterexample generator---that is, if a proof fails we can know why that happened. In this case we see \setlog code as formulas over the theory of finite sets (cf. Sect. \ref{prototypes}). In order to prove that a \setlog formula is a theorem we actually need to prove that its negation is unsatisfiable. For example, concerning \verb+rcvAddr+, we might prove that \verb+PsD+ and \verb+PsAs+ are disjoint sets. In this case we can submit the following to \setlog:
\begin{verbatim}
diff(Asm,As,D) &
PsD = ris(A in D,[],true,[this,A,connectMsg]) &
PsAs = ris(A in As,[],true,[this,A,addrMsg(As_)]) &
ndisj(PsD,PsAs).
\end{verbatim}
where \verb+ndisj(A,B)+ is interpreted as $A \cap B \neq \emptyset$. In this case \setlog returns \verb+no+ which means that the formula is unsatisfiable and so \verb+PsD+ and \verb+PsAs+ are disjoint.

As another example, we may want to prove that $s.acc$ is still a partial function after executing $CheckpointState$. This is a so-called \emph{state invariant preservation theorem}. These theorems are of the following form:
\begin{equation}\label{eq:invtheo}
Inv \land Op \implies Inv'
\end{equation}
that is, if $Inv$ holds before executing $Op$, then $Inv$ also holds in the after state (i.e., $Inv'$ is true, where $Inv' \defs Inv[\forall v, v'/v]$). If we want to use \setlog to discharge such a theorem we must check whether the negation of \eqref{eq:invtheo} is unsatisfiable. Then, in this case the formula to be submitted to \setlog is:
\begin{verbatim}
  World = {[acc,Acc] / Rest} &
  pfun(Acc) &
  checkpointState(World,Trans,World_) &
  World_ = {[acc,Acc_] / Rest} &
  npfun(Acc_).
\end{verbatim} 
where \verb+pfun(F)+ is interpreted as $F$ is a partial function and \verb+npfun+ as its negation. In this case \setlog returns \verb+no+, which means that \verb+checkpointState+ preserves the invariant, as otherwise expected.

\section{Final remarks}
\label{sec:conclusion}
Cryptocurrency software should be considered critical: its failures and vulnerabilities would cause irreparable loses. Formal methods have proved to be successful in delivering bug-free software for a range of critical systems. Hence, our first conclusion is that the cryptocurrency community should pay attention to formal methods. However, integrating formal methods into the development process of highly innovative domains cannot be done all at once. Then, our second conclusion is that the cryptocurrency community should \emph{gradually} adopt formal methods. 

Specifically, we propose to start the adoption process by using formal specification languages based on set theory and first-order logic. Most developers have been exposed to the mathematics underlying these languages, so adoption could be made minimizing the learning curve. Set-based specifications accurately and concisely describe cryptocurrency software. In order to support this claim we formally specified representative parts of two key components of cryptocurrency software. Once a set-based specification has been written, several verification techniques are enabled. In particular we have shown how prototypes can be easily generated and simulations can be run on them; then, we have applied a model-based testing method to generate test cases from a set-based specification; and finally, we have automatically discharged some proof obligations. 

However, these techniques do not allow the formal verification of the implementation. To this end more powerful languages and techniques (e.g. the Coq proof assistant \cite{coq-manual,coqart}) can and should be adopted in future stages. This should be done once the use of formal methods is well understood by the community.

Hopefully, the contents of this paper will convince key players of the cryptocurrency community to assess the application of formal methods more thoroughly. 

\bibliography{biblio}  

\newpage
\appendix

\section{\label{security}Subtleties and Traps with Security Properties}
Security properties have proved to be unexpectedly complex to grasp and formalize. One of the fundamental reasons for this is that security properties are not safety properties (w.r.t. the Alper-Schneider framework \cite{DBLP:journals/ipl/AlpernS85}). Most of the verification methods, results and tools developed for almost half a century by the Software Engineering community have been aimed at safety properties. Then, most of them fall short when dealing with security properties. Only very recently some of these methods and tools have been adapted for the problem of security properties.

For instance, safety properties enjoy, what is some times called, the \emph{refinement property}. This means that, if $S$ is the description of a system verifying safety property $F$ and $S'$ is a more deterministic refinement of $S$, then $S'$ (automatically) verifies $F$. However, in general, security properties do not enjoy the refinement property. This means that even if you have proved that $S$ verifies security property $F$ and that $S'$ is a refinement of $S$, then you cannot assert that $S'$ verifies $F$. The problem is that there are implementation decisions that in spite of generating a more deterministic system, they might introduce security issues. A typical example is the specification of the values stored in the memory cells delivered by the operating system to a process. At a certain level of abstraction these values might be underspecified which implies that, potentially, any value can be stored. Assume you can prove that this is secure. Afterwards, programmers decide to deliver the cells with the values left by the last process that used that buffer. Although this is an implementation of the first specification it is blatantly insecure.

On the other hand, reasoning about implementations provides the ultimate guarantee that
deployed mechanisms behave as expected. However, due to the issues mentioned above, formally proving non-trivial security properties of code might be an overwhelming task in terms
of the effort required, especially compared w.r.t. proving functional correctness (i.e. a safety property). In addition,
many implementation details are orthogonal to the security properties
to be established. This implies that slight changes in the implementation technology might have devastating consequences as concerns the security of the implementation. Therefore, complementary approaches are needed when non-trivial security properties are at stake:
\begin{enumerate}
\item Verification is performed on idealized models that abstract away the specifics of any particular implementation, and yet provide a realistic setting in which to explore the security issues that pertain to the realm of those (critical) mechanisms.
\item Additionally, verification is performed on more concrete models where low level mechanisms (such as pointer arithmetic) are specified. 
\item Finally, the low level model is proved to be a correct implementation of the idealized model.
\end{enumerate}

A particular class of the idealized models mentioned above are those called \emph{security models}. These models have played an important role in the design and evaluation of high assurance security systems. Their importance was already pointed out in the Anderson report~\cite{1296905}. The
paradigmatic Bell-LaPadula model~\cite{Bell:LaPadula}, conceived in 1973, constituted the first big effort on providing a formal setting in which to study and reason on confidentiality properties of data in time-sharing mainframe systems.

\end{document}